\documentclass[12pt]{article}
\usepackage{amssymb,bm,a4}

\newcommand{\eeq}{\end{equation}}
\newcommand{\beq}{\begin{equation}}
\newcommand{\ba}{\begin{array}}
\newcommand{\ea}{\end{array}}
\newcommand{\bea}{\begin{eqnarray}}
\newcommand{\eea}{\end{eqnarray}}

\newcommand{\vp}{\varphi}
\newcommand{\eps}{\epsilon}
\newcommand{\veps}{\varepsilon}

\newcommand{\mat}[1]{\mathbf{#1}}

\newcommand{\h}{{({\rm h})}}
\newcommand{\dem}{{({\rm d})}}
\newcommand{\CP}{CP}
\newcommand{\nd}{nd.\ }
\newcommand{\preprintno}[1]{\vspace{-2cm}{\normalsize\begin{flushright}#1\end{flushright}}\vspace{1cm}}

\begin{document}


\title{\preprintno{{\bf MCTP-02-30}} 
A realistic formulation of approximate \CP}

\author{Thomas Dent\thanks{email:\,tdent@umich.edu}\\
	{\em Michigan Center for Theoretical Physics, Randall Lab.,} \\
	{\em University of Michigan, Ann Arbor, MI 48109-1120} \vspace{3mm}\\
Joaquim Silva-Marcos\thanks{email:\,Joaquim.Silva-Marcos@cern.ch}\\
	{\em CERN, Geneva, Switzerland and 
IST$\,/\,$CFIF, Lisboa, Portugal}}


\date{\today}

\maketitle

\begin{abstract}
\noindent 
\CP\ violation in the SM is naturally implemented as a small imaginary 
perturbation to real Yukawa couplings. For example, a large \CP\ 
asymmetry in $B_d$ decays can arise if the imaginary parts of quark 
mass matrices are of order $10^{-3}m_{t,b}$ or smaller. Applying the 
principle of ``additive \CP\ violation'' to soft SUSY-breaking 
terms, the electric dipole moments of the neutron and mercury atom are 
predicted near current experimental limits; for nonuniversal $A$-terms, 
EDM bounds can be satisfied given certain flavour structures. 
The proposal may be formulated in a democratic basis, with Yukawas and 
soft terms of the form $(\mbox{const.})\times(1+\eps+i\zeta)$ where 
$|\eps|\ll 1$, $|\zeta| \lesssim 10^{-3}$, motivated by approximate 
permutation$\times\CP$ symmetry.
\end{abstract}



\newpage
\tableofcontents
\section{Introduction: Standard Model {\em vs.}\/ SUSY}\noindent
Recent measurements of a time-dependent \CP\ asymmetry in $B_d$ decays
\cite{BaBar,Belle},
in the context of the Standard Model, indicate a large unitarity triangle 
angle ($\sin 2\beta \simeq 0.7$), a correspondingly large CKM 
phase $\delta_{\rm KM}$, with a value of the Jarlskog invariant parameter
$J_{CP}=(3 \pm 1)\cdot 10^{-5}$. In contrast, the continued null
results of increasingly sensitive searches for fermion electric dipole 
moments (EDM's)
\cite{Harris:jx,electron,mercury},
in the context of softly-broken supersymmetry, strongly suggest that 
the complex phases of soft SUSY-breaking terms, namely gaugino masses, 
the Higgs bilinear $B$-term, and scalar trilinear $A$-terms, are of order 
$10^{-2}$ or smaller
\cite{SUSYEDM}; such bounds apply even in the limit of exact universality
of soft terms.

Thus supersymmetry appears to face a naturalness problem, since if, as 
usually imagined, $\delta_{\rm KM}$ arises from Yukawa phases of order
unity, one also expects large soft term phases. The alternative to small 
SUSY-breaking phases is soft terms with large CP-violating parts, but which 
cancel against each other in the expressions for EDM's. However, since 
recent experimental improvements provide three linearly independent 
limits, with a complicated dependence on the parameter space ruling out 
most of the parameter space where cancellation was claimed, this
possibility seems equally unnatural.
The ``SUSY \CP\ problem'' might also be circumvented by heavy (few TeV) 
scalar superpartners for the two light fermion generations
\cite{heavies},
which remains a possible solution, unless or until light superpartners are 
detected,\footnote{Although, see \cite{LebedevP,Chang} for recent 
challenges to the decoupling solution} or by assuming that SUSY-breaking 
takes certain special forms, for example gauge- or anomaly-mediation 
(which are, however, not without their own problems). 

An attractive and predictive alternative, requiring no assumptions about 
the SUSY-breaking sector, is approximate \CP\ symmetry
\cite{approximate}, 
usually formulated by requiring that all complex phases be small. 
Approximate \CP\ is motivated by spontaneous breaking of exact \CP\ symmetry 
\cite{CPgauge}, supposing that we live in a vacuum that happens to be close 
to a \CP-conserving one in the space of v.e.v.'s. The concept of 
approximate \CP\ thus relies on the existence of a measure of \CP\
violation in the theory, which will be important in the discussion. 
The proposal can be consistent with measurements in the $K^0$ system,
if there are supersymmetric contributions to \CP-odd flavour-changing 
interactions (which are required if $\delta_{\rm KM}$ is to be small). The
prediction for the $B_d$ decay asymmetry is small, hence approximate
\CP\ formulated in terms of small phases is ruled out (see also 
\cite{Babu:1999xf}).

If for some reason there are flavour-changing squark mass terms
with relatively large imaginary part, then $a_{J/\Psi K_S}$ can be 
generated by SUSY alone \cite{BrancoKhalil}; but it turns out that this 
possibility cannot be described as approximate \CP, since some phases 
in the soft breaking sector are not small (section \ref{sec:related}). 
Besides, with the increasingly exact fit of the SM unitarity triangle 
to \CP-violating observables, it becomes more difficult to see how 
supersymmetry can be the dominant contribution to such observables 
without some conspiracy between different soft terms.

We will argue that there is a viable alternative implementation of
approximate \CP, namely \CP\ violation with the {\em imaginary parts}\/ of
all couplings restricted to be small: we call this proposal ``additive
\CP\ violation''. It will be immediately objected that such imaginary
parts are not invariant under field redefinition. However, the usual
implementation via small phases is also subject to this ambiguity,
which reflects the well-known fact that the same physics may result 
from apparently different actions.

The proposal of approximate \CP\ --- either through small phases or small
imaginary parts --- requires a rule to deal with such
redefinitions. A particular set of coupling constants $\{\lambda\}$ is 
admissible as approximately \CP-symmetric if and only if it can be brought 
to a form $\{\lambda'\}$ by field redefinition, for which either the 
phases (in the usual implementation) or the imaginary parts (in our 
proposal) are of magnitude less than some given small number, and where the 
theory with ${\rm Im}\,\lambda=0$ would conserve \CP\ exactly. 
In other words, all phases or imaginary parts, respectively, larger than 
the given size must be removable (or reducible in size) by redefinition: 
non-removable phases or imaginary parts should be small.

This is logically distinct from the question whether the small phases 
or imaginary parts of a particular set of couplings $\{\lambda'\}$ can
be completely removed by field redefinition, in other words whether \CP\
is really violated. To obtain a reasonable phenomenology, small phases 
or imaginary parts must be non-removable. However it is usually somewhat 
laborious to find the field redefinition that reduces the size of such 
parameters to the absolute minimum, since flavour rotations must in general 
be considered. Thus we will not require always that small imaginary parts 
should be impossible to reduce by redefinition. Besides, the value of an 
experimentally observable measure such as $J_{CP}$ will quickly alert us 
to cases in which small imaginary parts can be made significantly smaller 
by field redefinition. 


We give examples in which both \CP\ violation and quark flavour can be 
satisfactorily described by adding small perturbations to an initial 
Lagrangian with unbroken \CP\ and flavour symmetry ({\em i.e.} rank
one fermion mass matrices and universal soft terms). In these
examples, ``small'' means of order $10^{-2}$ or less for dimensionless
quantities, or for mass terms, of order $10^{-2}$ or less compared with 
$v\approx 250\,$GeV, the natural mass scale of the SM and of softly-broken 
SUSY. Since the perturbation breaking \CP\ will mostly be of order $10^{-3}$, 
small soft term phases follow quite naturally in this proposal, and for 
universal soft terms EDM's are predicted below, but close to, current 
limits. We can also relax universality and allow the structure and size of 
\CP- and flavour symmetry-violating terms in the squark mass matrices to be 
comparable (but with coefficients differing by factors of order 1) to that 
in the quark mass matrices. Thus, while the smallness of the 
symmetry-violating terms remains to be explained, in this type of proposal
the SUSY-breaking terms do not appear unnatural in comparison to the 
(SUSY-preserving) Yukawa couplings.

\subsection{``Additive'' \CP\ violation}\noindent
There is currently no definite indication of the theoretical origin of \CP\ 
violation in Yukawa couplings and soft terms. The assumption that it occurs 
by complex phase rotations cannot give a good account of experimental 
results without an apparently unnatural distribution of phases between 
the Yukawas and soft terms. Both ``large'' and ``small'' phases are 
unsatisfactory, unless, as we shall see, ``small phases'' are implemented 
within a democratic model of flavour.\footnote{This exception also 
highlights the fact that the physical result of complex phases of a given 
size depends significantly on the choice of flavour basis.} 

We will explore the consequences if \CP\ violation originates from an 
imaginary perturbation to the couplings of an initial Lagrangian with real 
Yukawas and soft terms (section \ref{sec:imag}); the perturbations required 
will turn out to be small, usually order $10^{-3}$ or less, hence the 
proposal is an (unconventional) form of approximate \CP.

This formalism can also accommodate a theory of flavour: the total Lagrangian
of (potentially) \CP-violating and flavour-dependent operators is 
\beq \label{eq:additive}
\mathcal{L} = \mathcal{L}_0+\eps_1 \mathcal{L}_{f1}+ \eps_2\mathcal{L}_{f2} 
+\zeta \mathcal{L}_{\rm Im}
\eeq
where $\mathcal{L}_0$ has unbroken flavour symmetry (by which we mean rank 1
Yukawa matrices) and $\eps_1$, $\eps_2$ are small parameters generating
flavour structure, which are typically different in the up and down sectors,
and which we parameterise by $\eps_1\sim m_2/m_3$, $\eps_2\sim 
\sqrt{m_1m_2}/m_3)$, where $m_{1,2,3}$ are quark masses in ascending 
order. $\mathcal{L}_{\rm Im}$ then consists of Yukawas and soft terms with 
imaginary coefficients.\footnote{Note that $\mathcal{L}_{\rm Im}$ may not
in fact violate \CP, if it can be eliminated by field redefinition. 
Nevertheless, since we are only setting an {\em upper}\/ bound on $\zeta$,
the proposal remains well-defined.}

Alternatively, since $\zeta$ will turn out to be about the same size as 
$\eps_2$, the last two terms could be combined, with the interpretation that 
\CP\ is violated by small complex parameters, which also generate the small 
quark masses and mixing angles, but may also enter into non-flavour-dependent
couplings. In the case of spontaneously-broken \CP\ and flavour symmetry, 
Eq.~(\ref{eq:additive}) gives the effective couplings after integrating out 
the symmetry-breaking scalars, and $\eps_{1,2}$ and $\zeta$ are simply 
(products of) scalar v.e.v.'s normalised to some UV cutoff.

To take an extreme example, the quark mass matrices 
\bea \mat{m}^u&=&\frac{m_t}{3}\left( \ba{ccc} 
1 + i\xi & 1 & 0.9895 - i\xi \\
1 & 1 - i\xi & 0.9895 + i\xi \\
0.9895 - i\xi & 0.9895 + i\xi & 0.9944 \ea \right), \nonumber \\
\mat{m}^d&=&\frac{m_b}{3}\left( \ba{ccc} 
1.0844 & 1.0736 & 0.9312 \\
1.0736 & 1.0629 & 0.9527 \\
0.9312 & 0.9527 & 0.9377 \ea \right),
\eea
where $\xi=0.00033$, result in the CKM matrix 
\[ \left( \ba{ccc} 0.9748 & -0.2232+9\cdot 10^{-5}i & 
0.0008 - 0.0021 i \\ 0.2231 & 0.9740 & -0.0406 - 0.0005 i\\
0.0083 - 0.0022 i  & 0.03971 & 0.9992 \ea\right)
\]
(in the Wolfenstein phase convention) with a phase $\delta_{\rm KM}=-1.68$ 
and Jarlskog invariant $J=1.9\times 10^{-5}$. The complex phases of 
$m^{u,d}_{ij}$ are smaller than $7\times 10^{-4}$ and the range of 
nonremovable imaginary parts ({\em i.e.}\/\ 
Max$|{\rm Im}\,m_{ij}-{\rm Im}\,m_{ik}|$), which is independent of the 
flavour basis up to small numerical factors, is smaller than 
$2.5\times 10^{-4}m_t$.
 
Much of the paper is devoted to showing that acceptable values of 
$\delta_{\rm KM}$ and $J$ arise from equally small imaginary parts for Yukawa
couplings and that ``additive \CP\ violation'' can be consistent with 
correct masses and mixings (section \ref{sec:demo}). Then almost by
inspection, the small imaginary parts of soft terms that one naturally 
expects in this proposal satisfy the EDM bounds, at least in the limit of 
universality. Independently of any specific model of SUSY-breaking, for any 
soft mass term $\tilde{m}_i$ we have the expected form 
$\tilde{m}_i\sim ({\rm const.})\times (1+i\tilde{\zeta}_i)$, where 
$|\tilde{\zeta}_i|\lesssim 10^{-3}$ and \CP\ is conserved in the limit
$\zeta_i \rightarrow 0$. 

We also present a preliminary study of the proposal in the case of 
nonuniversal soft terms. Without a concrete model of SUSY-breaking we 
cannot make ``hard'' statements; however, it is very reasonable for the 
Lagrangian of flavour-dependent soft terms also to take a form analogous to
Eq.~(\ref{eq:additive}), and we proceed on this basis. With no restriction 
on the flavour structure of the imaginary perturbations Im$\,\tilde{m}_i$, 
the bounds on EDM's are exceeded by orders of magnitude. However, since 
we consider models of flavour structure in the Yukawas motivated by 
approximate symmetries, we have expectations for the flavour structure 
of nonuniversal $A$-terms, which give the main new contributions to EDM's. 
If the $A$-term matrices have a structure parallel to the Yukawa couplings, 
but with different (possibly complex) coefficients, EDM's are just at or 
slighly above experimental limits (section \ref{sec:nonU}). Off-diagonal 
imaginary parts of squark mass insertions are then very small, hence one 
does not expect SUSY contributions to flavour-changing interactions to be 
measurable.

These expectations are of course subject to modification, depending on
what concrete models of nonuniversal SUSY-breaking exist consistent with 
our proposal of ``additive \CP\ violation''. The main point of the paper is 
to give a rather general framework which, while being extremely simple 
to formulate, gives a realistic picture of \CP\ violation in both the 
SUSY-preserving and SUSY-breaking couplings of the MSSM. The proposal may 
result from, for example, spontaneous breaking of 
$P_L\times P_R\times CP$ symmetry by small v.e.v.'s, where $P_{L,R}$ are 
permutation symmetries acting on weak doublet and singlet matter 
generations respectively, and \CP\ symmetry enforces small imaginary 
parts for flavour singlet and flavour-dependent couplings alike. 

\subsection{Related work} \label{sec:related}\noindent
Of course, quark mass matrices consistent with Eq.~(\ref{eq:additive}) 
have been proposed previously, but little attention has been paid
to the manifest smallness of \CP\ violation in such {\em ans{\" a}tze}\/ 
and the implications for supersymmetry.

An exception is \cite{NirRattazzi}, in which \CP\ violation occurs 
through a small, complex parameter, that is constrained by a U$(1)^2$ 
flavour symmetry to appear only in certain off-diagonal elements in 
the quark masses and soft terms (to first order).
A large $\delta_{\rm KM}$ is produced through Im$\,m^d_{13}/m_b\sim 
10^{-2}$, while quark-squark alignment, coupled with the smallness of the 
\CP-violating parameter, sufficiently suppresses any SUSY contributions to 
the $K^0$ system and to EDM's. 

A solution to the SUSY \CP\ problem, based on the fact 
that \CP\ is only violated by flavour-changing interactions in the SM
(neglecting the QCD vacuum angle), was proposed in \cite{hermitian}. If 
the Yukawas and soft breaking terms and Hermitian in flavour space, 
``flavour-diagonal'' quantities are automatically real, while off-diagonal 
entries may be complex with {\em a priori}\/ arbitrary magnitudes and 
phases. EDM's are then suppressed, while allowing contributions to other 
\CP-violating observables from superpartner loops. However, Hermiticity 
does not arise from a field theory symmetry, except in left-right models 
\cite{Mohapatra_etal}, since it requires the exchange of the weak
doublet and singlet chiral fermions. Thus, the property is not preserved 
by radiative corrections (although the resulting effects on the EDM's are 
small). 

In \cite{BrancoKhalil} the authors studied the implications for 
supersymmetry of universal strength of Yukawa couplings (USY) 
\cite{Branco:fj}, a model of the democratic type which automatically 
has small phases and imaginary parts: the flavour structure is 
generated entirely by Yukawa phases. As a result of improved experimental
constraints since \cite{Branco:1995pw}, it is very likely impossible to 
obtain large $\delta_{\rm KM}$ and $J\gtrsim 10^{-5}$ with 
USY consistent with correct magnitudes of CKM elements \cite{Anna}.  
The authors chose a typical USY {\em ansatz}\/ with a much smaller value 
of $J$, thus all \CP-violating observables, including $a_{J/\Psi K_S}$, 
must be of supersymmetric origin.

Since they do not appreciably affect EDM's, one can take the off-diagonal 
squark mass terms in the super-CKM basis to have imaginary parts large 
enough to generate $a_{J/\Psi K_S}$. However, the real parts of the 
{\em same}\/ mass terms are tightly constrained by $B_d$--$\overline{B}_d$ 
mixing \cite{Becirevic}, so in this basis the term generating $a_{J/\Psi K_S}$ 
must have a {\em large}\/ phase. Even in the USY basis the authors chose 
some non-universal $A$-terms to have a phase $10^{-1}$; 
hence this proposal cannot be described as approximately \CP-symmetric. Thus 
the smallness of the phases of $B$, gaugino masses, and other $A$-terms 
(which are more tightly constrained by EDM's), still require 
some explanation.\footnote{However, it may be possible to reproduce the 
experimental value of $a_{J/\Psi K_S}$, with all phases in a particular 
basis being $\leq 10^{-2}$, using a more general democratic Yukawa 
ansatz in the presence of nonuniversal soft terms \cite{Shaaban}.}
In fact, as we show below, it is unusual for small phases in a democratic 
basis to produce small $\delta_{\rm KM}$ when the USY condition is relaxed: 
random small imaginary parts, of the size of the Im$\,y^{u,d}$ used in 
\cite{BrancoKhalil}, in most cases generate large $\delta_{\rm KM}$. 
The small values of $J$ and $\delta_{\rm KM}$ allow us to diagnose that the 
small imaginary parts considered by \cite{BrancoKhalil} may be reduced by 
some field redefinition; such a redefinition would, of course, 
result in Yukawas for which the USY condition was not manifest.

\section{Measures of \CP\ violation and small imaginary parts}\label{sec:imag}
\noindent
One cannot begin to solve the ``SUSY \CP\ problem'' without 
considering \CP\ violation in the SM, {\em i.e.}\/\ in the CKM matrix 
$\mat{V}$, which arises in changing from the weak interaction basis of 
quarks to the mass eigenstate basis. Both \CP\ violation and flavour 
originate from the quark mass matrices, and thus from Yukawa couplings 
$y^{u,d}_{ij}$. Our understanding of \CP\ violation depends on 
what assumptions are made about the form of the mass matrices and where 
their phases or imaginary parts come from. 

In the SM, one cannot construct a \CP-violating observable without 
involving all three generations and bringing into play the small (13) and 
(23) elements of $\mat{V}$. Any such quantity is highly suppressed, 
compared to, say, a charged current amplitude involving diagonal or 
(12) elements of $\mat{V}$: thus {\em any \CP-violating effect in the SM
involves small amplitudes}, where ``small'' means suppressed by at least 
three orders of magnitude. The ``smallness'' of \CP\ violation is manifested 
in the Jarlskog parameter, and to some extent in the $K^0$ system. In the 
$B_d$ system, a large \CP\ asymmetry in a particular channel simply means 
that we are comparing with a \CP-even quantity that also happens to be 
very small: the branching fraction into any channel with a large \CP\ 
asymmetry is inevitably suppressed. The same comparison also leads to 
$\delta_{KM}$ being order 1: we will argue that $\delta_{KM}$ is not a 
good measure of \CP\ violation in most circumstances.

The usual picture of \CP\ violation is to start with Yukawas 
$y^{u,d}_{ij}$ in a ``heavy'' or hierarchical basis (for which the $(33)$ 
elements are large and the rest small), and introduce large (order 1) 
phases for some or all elements. In this picture, which can be 
generalised to superpartner interactions defined in the same flavour 
basis, \CP\ violation appears to be a large effect. However, such 
phases may not be a good measure of \CP\ violation on the space of 
couplings, as we discuss later: for the moment, note that large phases
can always be removed from the large Yukawa couplings by field 
redefinitions, and also can change by orders of magnitude under a 
(real orthogonal) change of flavour basis, even after the phases of the
large Yukawas have been removed). 

In any given flavour basis, fields can be redefined by phases to reduce the 
size of Im$\,y_{ij}$ as far as possible. The remaining imaginary parts 
Im$\,y^{u,d}_{ij}$ 
stay about the same size under change of basis: thus they are largely 
independent of which theory of flavour one considers. Hence we think they
are a better candidate for a measure of \CP\ violation.
In the SM, such imaginary parts can be smaller than $10^{-3}$ and still 
be consistent with $\delta_{\rm KM}\sim 1$. If one similarly redefines 
phases on $\mat{V}$ to reduce imaginary parts, one finds the Wolfenstein 
form with Im$\,V=\,$few$\,\times 10^{-3}$. However, the relation 
between Im$\,\mat{y}^{u,d}$ and Im$\,\mat{V}$ is more subtle and depends 
to some extent on flavour structure, as we discuss later. 


\subsection{``Amplification'' of $\delta_{\rm KM}$}
A large KM phase results rather generically, if all quark 
Yukawa couplings (normalised to the largest coupling in the up or down 
sector respectively) have imaginary parts less than or equal to $10^{-3}$; 
equivalently, $|{\rm Im}\,m^{u,d}_{ij}|\leq 10^{-3}m_{t,b}$.
\footnote{We normalise Im$\,m^{d}_{ij}$ to $m_b$ because the bottom Yukawa
could be order 1 if $\tan\beta_H \equiv \langle H_U\rangle/ 
\langle H_D\rangle$ is large; if $\tan\beta_H\simeq 2$ then we might in 
principle allow Im$\,m^{d}_{ij}/m_b\simeq 0.05$, since then we would have
Im$\,m^{d}_{ij}/m_t\sim 10^{-3}$. We take the more conservative limit on 
Im$\,m^{d}$.} 
In a democratic basis, all mass matrix elements are equal to 
$\frac{1}{3}m_{t,b}$ up to small perturbations, and we can have 
$|{\rm Im}\,m^{u,d}_{ij}|/m_{t,b}\simeq 3\times 10^{-4}$, thus the 
(relative) Yukawa phases Arg$\,y_{ij}$ are $10^{-3}$ or smaller. The 
democratic basis has the obvious advantage that small phases and small 
imaginary parts mean the same thing, therefore ``additive \CP\ violation'' 
can be formulated unambiguously in this basis. A similar type of mass 
matrix, except that the imaginary parts Im$\,\mat{m}^d/m_b$ were somewhat 
larger, was described in \cite{SilvaM01}.

Given this {\em ansatz}, the imaginary parts of the diagonalisation matrices 
$\mat{U}^{u,d}_L$ will be larger: for random distributions of small 
imaginary parts, we find 
$|{\rm Im}\,\mat{U}^{u}_L|\lesssim 5\times 10^{-2}$ and 
$|{\rm Im}\,\mat{U}^{d}_L|\lesssim 5\times 10^{-3}$. The 
{\em amplification}\/ of Im$\,\mat{U}^{u,d}$~\footnote{We suppress the 
$L$ suffix unless there is ambiguity.} relative to 
Im$\,\mat{m}^{u,d}/m_{t,b}$ is an essential part of our argument, and 
is related to (the inverse of) the small parameters that generate 
quark masses and mixings. 

The complex phases of $U^{u,d}_{ij}$ are also not necessarily large: 
for mass matrices near the democratic form these matrices have elements 
of order 1. It is straightforward to see how (small) imaginary parts of 
$\mat{U}^{u,d}$ feed into a realistic CKM matrix (section 
\ref{sec:makingJ}). 

In order to obtain $J= (3\pm 1)\times 10^{-5}$, we find that 
Im$\,\mat{U}^{u,d} \simeq 3\times 10^{-3}$ is the absolute minimum; but 
one requires very specific structures of Im$\,\mat{U}^{u,d}$, and of 
the mass matrices, for $J$ to be generated so ``efficiently''. In most 
cases we find $J/{\rm Im}\,\mat{U} < 8\times 10^{-4}$, implying that 
Im$\,\mat{U}^u$ or Im$\,\mat{U}^d$ should be order $5\times 10^{-2}$, 
as found above. However, the flavour structure of small imaginary 
parts cannot be random, since it is constrained by the requirement 
that the quark mass hierarchy and mixing angles be stable, particularly 
the up mass and $|V_{13}|$. Even after imposing this requirement, several 
possibilities remain: see section \ref{sec:demo}.


\subsection{Phases {\em vs.}\/\ imaginary parts}
Many \CP-violating observables do not directly determine the complex phases 
of (physical or invariant combinations of) 
couplings, but only the imaginary parts. For the soft terms, one does 
not in general know either the real part or absolute magnitude, whereas 
for the KM phase or $\sin 2\beta$, both the magnitude and imaginary part
of the relevant quantity can be determined, so one may convert to a 
complex phase. We argue that imaginary parts of (physical, 
rephasing-invariant) soft terms, and 
$J=\pm{\rm Im}\,V_{ij}V_{kl}V_{il}^*V_{kj}^*$ for the CKM matrix, are 
suitable quantities to compare to experiment and to characterise the 
strength of \CP\ violation theoretically. If we normalise symmetry-violating 
effects by comparison to symmetry-preserving ones, which is the hidden 
assumption behind using complex phases, we get very different answers 
depending on which \CP-even quantity we choose. The resulting measures are 
somewhat arbitrary: the size of \CP-even quantities does not tell us much 
about \CP\ violation.

The quantity $\tilde{\delta}^u\equiv (\delta^u_{11})_{LR}M_3/\tilde{m} = 
v\hat{A}^{u*}_{11}M_3/\tilde{m}^3$, the $(11)$ mass matrix element mixing L 
and R squarks in the super-CKM basis, times the gluino mass, 
normalised to an average squark mass, is dimensionless and rephasing 
invariant, and directly enters diagrams generating an EDM for the neutron. 
From experimental bounds, $|{\rm Im}\,(\delta^u_{11})_{LR}
M_3|/\tilde{m}$ should be less than about $10^{-6}$. From this bound,
{\em almost nothing}\/ can be deduced about the phase 
Arg$\,\tilde{\delta}^u$ without further assumptions. One might expect 
Re$\,\tilde{\delta}^u$ to be of order $m_u/\tilde{m}\sim 10^{-5}$, but 
this can be highly model-dependent. With non-universal $A$-terms the real 
part might be much larger, implying a very strict bound on the phase, or 
it might even be smaller, given some (rather bizarre) structure of soft 
terms: neither is as yet experimentally excluded.\footnote{Bounds on 
Re$\,\tilde{\delta}^u$ exist but are much weaker than $10^{-5}$
\cite{Gabbiani96}.} The size of soft term phases is thus an ill-defined 
way to discuss the SUSY \CP\ problem, unless one takes some model-dependent 
assumption such as minimal supergravity. For the $B$/$\mu$ contributions, 
which lead to a limit on Im$\,\mu M_{1,2}/\tilde{m}^2$ (in the phase 
convention where $B\mu$ is set real), limits on the phase are somewhat 
better-defined since $|\mu|$ and $|B\mu|$ are is constrained by correct 
electroweak symmetry-breaking, but the experimental limit is still found 
more directly for the imaginary part.

Now we give some examples showing that $\delta_{\rm KM}$ and $(\sin2)\beta$ 
are not sensible measures of \CP\ violation over the space of possible
values of the CKM matrix. In the Standard Model, consider the limit 
$|V_{13}|\rightarrow 0$, $\delta_{\rm KM}={\rm const.}$, achieved by taking 
$\theta_{13}\rightarrow 0$ with all other angles constant. Clearly
Im$\,\mat{V}$ and $J$ vanish in this limit, and all \CP-violating 
effects become unmeasurably small. Nevertheless, by the standard lore, 
\CP\ violation would remain ``large''! Approaching this limit, the 
unitarity triangle would be the same shape, the time-dependent decay 
asymmetry $a_{J/\Psi K_S}$ might well remain order 1, but eventually these 
measures would become meaningless, since measurements could not be made.

Now consider the case of $|V_{13}|$ becoming larger by a factor of 2, 
with Im$\,V_{13}$ varying such that $J$ is constant (and other elements 
adjusted to preserve unitarity).
\CP-violating signals such as the EDM's and $\eps_K$ would stay the same 
size, but the conventional measures $\delta_{\rm KM}$ and $\sin 2\beta$ 
would become smaller. In the case of a \CP-odd rate asymmetry defined as 
$(R-\overline{R})/(R+\overline{R})$ for some rare decay channel, the 
difference $R-\overline{R}$ which signals \CP\ violation would not change 
for a given luminosity, but the total rate $R+\overline{R}$ might be larger. 
Division by $(R+\overline{R})$ is convenient, but the resulting ratio does 
not tell us anything essential about the size of \CP\ violation. 
One might equally well use $(R-\overline{R})/R_{\rm tot}$ as a measure of 
symmetry violation (where $R_{\rm tot}$ is the total rate over all 
channels), which would likely stay constant (and small) with constant 
$J$. In both these cases, $\delta_{\rm KM}$ and $\sin 2\beta$ give a 
misleading answer to the question of how large the symmetry violation is. 

In what follows we will evaluate
both $\delta_{\rm KM}$ and $J$, but one should keep in mind that 
$\delta_{\rm KM}$ is a somewhat meaningless measure unless the quark 
masses and mixing angles are sufficiently close to the observed values.


\section{Small imaginary parts in democratic {\em ans{\" a}tze}}
\label{sec:demo}
\noindent
In this section we study the influence of adding small imaginary parts or 
complex phases to quark mass matrices which are initially real and close 
to the democratic form $\mat{\Delta}$ with all entries equal to 1. 
As noted before, small imaginary parts and small phases are equivalent in 
this basis. The flavour structures we use are consistent with successive 
breaking of a symmetry group permuting 3 generations, to the 2-element 
group, then to nothing.

The initial structure for the quark mass matrices is
\beq \label{eq:initial}
\mat{M}^{u,d}_0 \equiv \frac{3\mat{m}^{u,d}_0}{m_{t,b}} 
= \mat{\Delta} + \eps \mat{D}_0,
\eeq
where $\mat{m}^{u,d}$ is the conventionally normalised quark mass matrix,
$\Delta$ is the ``democratic'' matrix with all entries equal to unity, 
$\eps$ is a small parameter and $\mat{D}_0$ is a real matrix of
order $1$. When one adds small imaginary parts to $\mat{M}_0$ to obtain 
\beq \label{eq:epszeta}
\mat{M}_\zeta = \mat{\Delta} + \eps\mat{D}_0+ i\zeta\mat{D}/2, 
\eeq
this will in general violate \CP\ and also change the mass spectrum and 
mixing angles. Note that the imaginary part is normalised so that the 
largest {\em relative}\/ phases are of order $\zeta$, the entries of 
$\mat{D}$ taking both signs.

The up quark mass is most likely to receive a significant contribution 
from the imaginary part, being the smallest eigenvalue of the initial 
mass matrices $\mat{M}_0^{u,d}$. 
Similarly, $V_{ub}\equiv V_{13}$, being the smallest
accurately-measured CKM matrix element, is likely the most sensitive to
imaginary perturbations (although $V_{12}$ may also be sensitive since it
usually depends on small (differences between) mass matrix elements). 
If we assume for example that $\mat{M}_0^u$ has a 
vanishing smallest eigenvalue and that $\zeta < \eps$, then it is clear 
that, unless $\mat{D}$ has some special structure, 
$\det \mat{M}^u_\zeta$ will be proportional to $\zeta \eps/2$. 
A quick calculation also yields that $\chi$, the second invariant of 
the Hermitian matrix $\mat{H}\equiv\mat{M}\mat{M}^\dag$, receives 
contributions proportional to $4\eps^2$. The largest eigenvalue of 
$\mat{M}^u_\zeta$ is equal to $3$ to a good approximation and the 2\nd 
eigenvalue is much smaller, thus the smallest eigenvalue is given at 
leading order by 
$3m_u/m_t \sim \det\mat{M}_\zeta/\sqrt{\chi} \sim \zeta/4$.
Thus, for random $\mat{D}$ the largest value of $\zeta$ that one can 
allow in the up sector is about $12m_u/m_t\sim 1.2\times 10^{-4}$, a 
severe constraint which as we will see prevents a large value of 
$\delta_{\rm KM}$. One can try to evade this by choosing an initial 
$\mat{M}_0^u$ which has a smallest eigenvalue different from zero, but if
$\zeta$ becomes too large the conclusion is unavoidable. The same arguments 
apply to the down sector, the largest allowable relative phase being of 
order $1.4\times 10^{-2}$.

\subsection{Two democratic schemes}\noindent
As an example, take the democratic {\em ansatz}\/
\begin{equation} \label{eq:dem1}
\mat{M}^{u,d}_0 = 
\left( \ba{lll}
1 & 1 & 1+b \\
1 & 1 & 1+b-c \\
1+b & 1+b-c & 1+b-c
\ea \right)
\end{equation}
where $b=9m_2/2m_3$, $c=\pm 3\sqrt{3m_1m_2}/m_3$. 
These mass matrices reproduce the observed quark masses\,\footnote{Note that $m_{1,2,3}$ are input parameters and will be only 
approximately equal to the resulting mass eigenvalues.}
and mixings reasonably well, with remarkably few free parameters 
($c_u$ and $c_d$ taking $(+)$ and $(-)$ signs respectively). 

We also consider the mass matrices 
\bea \label{eq:demmass}
\mat{m}^{u,d}_0=m^{t,b} \left[ \frac{1}{3}
\left( \ba{ccc} 1& 1& 1\\ 1& 1& 1\\ 1& 1& 1\ea\right) +\frac{B}{3\sqrt{2}}
\left( \ba{ccc} 2& 2& -1\\ 2& 2& -1\\ -1& -1& -4\ea\right) +\right.\nonumber 
\\ 
\left.+\frac{C}{6} \left( \ba{ccc} 1& 1& -2\\ 1& 1& -2\\ -2& -2& 4\ea\right)
+\frac{D}{\sqrt{3}} \left( \ba{ccc} 1& 0& -1\\ 0& -1& 1\\ -1& 1& 0\ea
\right) \right]
\eea
where $B=1(0.9)m_1/m_2$, $C=m_2/m_3$, $D=1.1(1.3)\sqrt{m_1m_2}/m_3$ 
in the up (down) sector. With all entries real, and slightly
different input parameters $m_i$, this also gives acceptable masses and
mixing angles \cite{demmass}.

To introduce \CP\ violation, we keep the same magnitudes $|m_{ij}|$ 
but introduce phases $\zeta^{u,d}_{ij}$ where $|\zeta^{u,d}_{ij}|\leq 
10^{-2}$: this {\em ansatz} is clearly of the form (\ref{eq:initial}). 

\subsection{Random small imaginary perturbations}\label{sec:random}\noindent
In the absence of definite clues as to the origin of \CP\ violation, one
can imagine the phases $\zeta$ in the democratic basis to be random 
subject to the above constraint: $\zeta^{u,d}_{ij}=\zeta d^{u,d}_{ij}/2$ 
where $d^{u,d}_{ij}$ takes a uniform distribution on $(-1,1)$. 
Then for each initial mass matrix and random set of $d_{ij}$,
we calculated quark masses, mixings and \CP\ violating parameters as 
functions of $\zeta$. For small enough $\zeta$ one expects a linear 
behaviour of $\delta_{\rm KM}$ and $J$ and a quadratic variation of the 
masses away from their values at $\zeta=0$.
\footnote{A imaginary, Hermitian perturbation of the real mass-squared 
matrix $\mat{H}_0\equiv \mat{M}_0\mat{M}^{\dag}_0=\mat{M}_0\mat{M}^T_0$ 
leaves the eigenvalues unchanged to first order; the proof is simple.}

To emphasize the point that small imaginary parts, or small phases in 
the democratic basis, lead to large $\delta_{\rm KM}$, we define an 
``amplification factor'' $F_{CP}\equiv \delta_{\rm KM}/\zeta$.
is the largest {\em relative}\/ Yukawa phase allowed. 
We expect $F_{CP}$ to tend to a constant in the limit of small $\zeta$, and 
it turns out that the asymptotic value is in most cases of order 
$m_{n+1}/m_{n}$ or $\hat{\theta}^{-1}$, and may be larger. (We give an 
analytic derivation of the related quantity $J/\zeta$ in section 
\ref{sec:makingJ}.)
With many uncorrelated small imaginary perturbations 
$\zeta_{ij}\equiv\zeta d_{ij}$, the contributions to $F_{\CP}$ add 
(although accidental cancellations are possible) and the largest 
amplification factors, of order $10^3$, win. 

We display two typical sets of results for each mass {\em ansatz}. For
the two-parameter matrices Eq.\ (\ref{eq:dem1}),
\begin{table}
\caption{Dependence of CKM parameters and $m_u$ on $\zeta$ for the
two-parameter mass matrices, for two sets of random coefficients 
$d^{u,d}_{ij}$.\label{random1}}
\centering
\begin{tabular}{|c|ccccc|} \hline
$\zeta$& $\delta_{\rm KM}$& $F_{CP}\equiv\delta_{\rm KM}/\zeta$& $J\times 10^5$& 
$m_u(\mbox{MeV})$& $|V_{13}|\times 10^3$ \\ \hline
$0.0001$ 

& $-0.065$, $0.033$ 
& $-650$, $340$ 
& $0.16$, $-0.079$ 
& $4.0$, $4.1$ 
& $2.2$, $2.2$ 
\\

$0.0003$ 

& $-0.19$, $0.100$ 
& $-645$, $330$ 
& $0.47$, $-0.24$ 
& $6.0$, $7.0$ 
& $2.3$, $2.2$ 
\\

$0.001$ 

& $-0.574$, $0.32$ 
& $-570$, $330$
& $1.56$, $-0.79$ 
& $17$, $20$ 
& $2.7$, $2.3$ 
\\

$0.003$ 

& $-1.06$, $0.79$ 
& $-350$, $260$ 
& $4.6$, $-2.3$ 
& $49$, $61$
& $5.0$, $3.0$ 
\\ 

$0.01$ 

& $-1.25$, $1.32$ 
& $-125$, $130$ 
& $13.6$, $-6.6$ 
& $177$, $230$ 
& $14$, $6.3$ 
\\
\hline
\end{tabular}
\end{table}
we find large amplification factors in the CKM phase, such that it reaches 
order 1, and the Jarlskog parameter is the correct order of magnitude, 
for random phases of order $0.003$ (Table \ref{random1}). The amplification 
becomes smaller when $\delta_{\rm KM}$ is no longer small, and enters a 
nonlinear regime, although $J$ continues to grow in a more nearly linear 
fashion. The CKM mixing angles and masses do not vary greatly with 
$\zeta$, except for the up mass and $V_{13}$. Already for 
phases less than $0.0003$ the up mass is affected the perturbation. 
Recall that the upper bound on the largest relative phase is $\zeta 
\leq 12m_u/m_t\simeq 6\times 10^{-5}$ if the up mass is not to be 
affected. Also, having been constant at small phases, $|V_{13}|$ receives 
contributions which result essentially from adding an imaginary part 
of magnitude similar or greater to the initial real part $V_{13}|_0$.
Since for $\delta_{\rm KM}$ to be order 1 we only require the imaginary 
part to be about the same size as $V_{13}|_0$, the absolute value 
can increase by a factor of at most two in reaching the correct value 
of $J$, which is not problematic and can actually give a better fit 
to experiment than in the case of no \CP\ violation.

For the three-parameter mass matrices, Table \ref{random2},
\begin{table} 
\caption{Dependence of CKM parameters and $m_u$ on $\zeta$ for the
three-parameter mass matrices, for two sets of random coefficients 
$d^{u,d}_{ij}$.\label{random2}}
\centering
\begin{tabular}{|c|ccccc|} \hline
$\zeta$ & $\delta_{\rm KM}$ & $F_{CP}\equiv\delta_{\rm KM}/\zeta$ & $J\times 10^5$ & 
$m_u(\mbox{MeV})$ & $|V_{13}|\times 10^3$ \\ \hline
0.0001

&$-0.122$, $0.047$ 
&$-1220$, $470$ 
&$0.27$, $-0.10$ 
&$4.3$, $5.7$ 
&$2.1$, $2.1$ 
\\

0.0003

&$-0.35$, $0.14$ 
&$-1170$, $470$ 
&$0.81$, $-0.31$ 
&$4.8$, $12.3$ 
&$2.2$, $2.1$ 
\\

0.001

&$-0.86$, $0.45$ 
&$-860$, $450$ 
&$2.7$, $-1.0$ 
&$8.6$, $38$ 
&$3.4$, $2.3$ 
\\

0.003

&$-1.17$, $0.98$ 
&$-390$, $330$ 
&$7.8$, $-3.1$ 
&$23$, $115$ 
&$8.1$, $3.6$ 
\\

0.01

&$-1.01$, $1.47$ 
&$-101$, $148$ 
&$19$, $-10$ 
&$89$, $376$ 
&$23$, $9.9$ 

\\
\hline
\end{tabular}
\end{table}
the amplification may be still larger, such that one achieves acceptable
values of $\delta_{\rm KM}$ and $J$ for $\zeta=10^{-3}$, but the value of 
$m_u$ is still unacceptably high for these parameter values.
We also see that $|V_{13}|$ grows too large with larger values
of $\zeta$. In fact, in both {\em ans{\" a}tze}\/ we found that $|V_{12}|$
and $|V_{21}|$ could also exceed the experimental bounds for larger values 
of $\zeta$, although not so drastically as $|V_{13}|$. This can be traced
to the fact that the (1-2) mixing also originates from entries of order
$\sqrt{m_1m_2}/m_3$ which are sensitive to small perturbations.

The instability of the up mass shows that as expected, a random structure 
of imaginary parts is strongly disfavoured to produce the observed \CP\ 
violation.
If one enforces $\zeta^u=0$ for the up-type phases, then the problem is 
reduced to keeping $m_d$ stable; but in this case the amplification factor 
$F_{CP}$ tends to be much smaller, of order $50$ or less, such that much 
larger phases $\zeta$ are required to produce $\delta_{\rm KM}\sim 1$. The 
down mass is then marginally unstable, and the size of the required phases 
or imaginary parts is only marginally consistent with \CP\ violation being a 
small perturbation.

Thus, any small perturbation giving rise to realistic \CP\ violation 
must have a more specific structure, which should in general be 
correlated with the structure of the original real mass matrices. 
In the next section we show how the amplification of small imaginary parts 
occurs, then in section \ref{sec:struct} give some structures of quark mass 
matrices that preserve a correct mass hierarchy under imaginary 
perturbations consistent with $\delta_{\rm KM}$ being large and $J$ of 
order few$\times 10^{-5}$.

\section{How amplification works}
\subsection{From Im$\,\mat{U}^{u,d}$ to $J$}\label{sec:makingJ}\noindent 
For diagonalisation matrices $\mat{U}^{u,d}$ for which large
imaginary parts have been removed by field redefinition, leading 
contributions to $J$ come from small imaginary parts Im$\,U^u_{ij}$ 
multiplying large real parts of $\mat{U}^d$, and vice versa.
Consider, in a ``heavy'' or hierarchical basis, small imaginary 
perturbations to the $\mat{U}^{u,d\h}$ matrices, which correspond to 
infinitesimal U$(3)$ transformations with symmetric generators.\footnote{Written as $\mat{U}_\omega = 1+i\omega_i\mat{S}_i$ with real symmetric $\mat{S}$.}
For the purpose of estimating $J$ we write 
$\mat{V} = \mat{U}_0^{u\h T}\mat{U}_\omega\mat{U}^{d\h}_0$, where 
$\mat{U}^{u,d}_0$ are real. Then we find that imaginary parts in 
$\mat{U}^{u,d\h}$ that can be expressed as diagonal $\mat{U}_\omega$ give 
negligibly small contributions to $J$ (since off-diagonal elements of 
$\mat{U}^u$ are very small). 
With a small imaginary part $\omega_{12}$ in the (1,2) and (2,1) elements 
of $\mat{U}_\omega$ we find $J/\omega_{12} = \lambda^5 A^2(1-\rho)\simeq 
3\times 10^{-4}$ in terms of Wolfenstein parameters; for small
$\omega_{23}$ and $\omega_{13}$ defined analogously we find 
$J/\omega_{23}=-A\lambda^4\rho \simeq -4\times 10^{-4}$, 
$J/\omega_{13} = A\lambda^3 \simeq 9\times 
10^{-3}$ respectively. Thus we require either $\omega_{12}\sim 0.1$, 
$\omega_{23}\sim (-)1/13$ or $\omega_{13}\sim 1/300$ for $J$ to take the
correct experimental value. 

\subsection{Diagonalisation}\label{sec:diag}\noindent
To see how Im$\,\mat{U}$ can be much larger than Im$\,\mat{m}/m_3$
and thus how the observed values of $J$, $\delta_{\rm KM}$ can result from
small imaginary parts, consider the mass matrix
\begin{equation}
\mat{m}^{\h} = m_0 
\left( \ba{ccc} 
0 & w +i\zeta^\h_{12} & i\zeta^\h_{13} \\ 
w +i\zeta^\h_{12} & v +i\zeta^\h_{22} & u +i\zeta^\h_{23} \\
i\zeta^\h_{13} & u +i\zeta^\h_{23} & 1+i\zeta^\h_{33}
\ea \right)
\end{equation}
where $u$, $v\sim m_2/m_3$, $w\sim \sqrt{m_1m_2/m_3}\sim(m_2/m_3)^{3/2}$
and the $\zeta^\h_{ij}$ are of order $c$. Such a mass matrix in the 
``heavy'' basis can always be related to a democratic one of the type
we are considering via
\beq \label{eq:mhier}
\mat{m}^{\h}\equiv \mat{F}\mat{m}^{\dem}\mat{F}^T 
\eeq
where 
$\mat{F} = \frac{1}{\sqrt{6}} {\small
\left(\!\ba{ccc} 
\!\sqrt{3}& -\sqrt{3}& 0\!\\ \!1& 1& -2\!\\ \!\sqrt{2}& \sqrt{2}& \sqrt{2}\! 
\ea\! \right) }$ 
diagonalises $\mat{\Delta}$; we do the analysis in the heavy basis since it
is somewhat simpler. We consider symmetric matrices for simplicity
and do not give a perturbation to $m_{11}$ since the smallest mass 
eigenstate would then be unacceptably unstable.

Then we construct $\mat{H}\equiv \mat{M}\mat{M}^{\dag}$ 
which is diagonalised as $\mat{U}^\dag_L \mat{H} \mat{U}_L= 
\mbox{diag}(m_i^2)$. The eigenvalues are given by 
$m_0(1+\mathcal{O}(u^2),\, v^2+\mathcal{O}(\zeta^{\h2}),\,
\mathcal{O}(w^4,\zeta^{\h4})/v^2)$ and we find that $\mat{U}_L$ is 
given by 
\beq
\mat{U}_L \simeq
\left(\ba{ccc} 
\! 1 & (w+i\zeta^\h_{12})/v & uw+i \zeta^\h_{13} \\ 
\! (-w+i\zeta^\h_{12})/v & 1 & u+i\zeta^\h_{23} \\ 
\! (uw+i(v\zeta^\h_{13}-u\zeta^\h_{12}))/v & -u+i\zeta^\h_{23} & 1
\ea\right)
\eeq
where we impose that the diagonal elements be real and keep only the 
leading real and imaginary parts (thus the matrix is only approximately 
unitary). Clearly the $(12)$ imaginary part is amplified by the 
diagonalisation; the diagonal perturbations $\zeta^\h_{ii}$ disappear 
(being unphysical as far as \CP\ violation is concerned), while the $(23)$ 
and $(13)$ perturbations are unchanged in size. Identifying the imaginary 
parts of $\mat{U}_L$ with the parameters $\omega_{12}$, {\em etc.}\/, of 
the previous discussion, we have
\bea
\frac{J}{\zeta^\h_{12}} &\sim& \frac{\lambda^5 A^2(1-\rho)}{v} \simeq 0.07\
({\rm up}),\ 0.006\ ({\rm down}), \nonumber \\
\frac{J}{\zeta^\h_{23}} &\sim& -A\lambda^4\rho \simeq 
-4\times 10^{-4},\mbox{ or} \nonumber \\
\frac{J}{\zeta^\h_{13}} &\sim& A\lambda^3 \simeq 9\times 10^{-3}
\eea
where each perturbation is considered separately.
If we assume that the observed value $J \simeq 3\times 10^{-5}$ 
is correlated with $\delta_{\rm KM}= 0.5$--$1$, we estimate the total 
amplification to be
\beq
\frac{\delta_{\rm KM}}{\zeta^\h_{12}}
\lesssim \frac{J}{\zeta^\h_{12} J_{\rm exp}} \sim 2300 ({\rm up}),\ 
200 ({\rm down}) 
\eeq
since, other things being equal, the largest contribution evidently comes
from $\zeta^\h_{12}$; subleading contributions are from
\beq
\frac{\delta J}{\zeta^\h_{13}J_{\rm exp}} \sim 300,\ 
\frac{\delta' J}{\zeta^\h_{23}J_{\rm exp}} \sim 13. \nonumber
\eeq
Recall that the imaginary parts of $\mat{m}^{u,d}$ in the democratic 
basis are of order $\zeta^{u,d}_{ij}m_3/3 =\zeta |d^{u,d}_{ij}|m_3/3$,
thus if $d^{u,d}_{ij}$ are random in this basis one expects 
$\delta_{\rm KM}$ to be $\sim 750\zeta$ (or $\sim 65\zeta$, if only the 
down sector contributes), depending on the relative signs and magnitudes
of the entries that feed into $\zeta_{12}^\h$. Thus the numerical results
for the dependence of $\delta_{KM}$ and $J$ on $\zeta$ can be understood 
analytically, at least in the linear regime.

\section{Special mass matrices and phase structures}\label{sec:struct}
\noindent
In order to preserve the small mass eigenvalues, we require $\det\mat{M}$
to remain of order $m_1m_2/m_3^2$ or $(m_n/m_{n+1})^3\sim \delta_0^3$ under 
the \CP-violating perturbation. As we argued previously, the expectation 
in the absence of any special structure is for 
$\det\mat{M}\sim\veps\delta_0$; for phases of order $\delta_0$, which are 
likely required to generate large $\delta_{\rm KM}$, the lightest mass 
eigenvalue would then approach the same magnitude as the 2\nd eigenvalue. 

\subsection{Decoupling of quark mass and \CP}\label{decoup} 
\noindent
The most simple way to introduce \CP\ violation in an almost democratic 
structure without spoiling the mass spectrum is of course one that
leaves the masses exactly invariant. This is possible just by 
left multiplying $\mat{M}_0$ with pure phase matrices of type 
$\mat{K}=\exp(i\zeta \cdot{\rm diag}(\vp_1,\vp_2,\vp_3))$ 
with $\zeta$ small.\footnote{Right multiplication would of course be unobservable and unphysical, corresponding to field redefinitions.} 
The mass hierarchy is then decoupled from \CP\ violation, suggesting that 
the two effects have different origin. Given
\beq \label{eq:mk}
\mat{M}^u=\mat{M}_0^u,\ \mat{M}^d=\mat{K} \mat{M}_0^d
\eeq
it is easy to calculate the CKM matrix as $\mat{V}=\mat{U}_0^{u\dag} 
\mat{K} \mat{U}_0^d$. Clearly, introducing another diagonal matrix of 
phases for the up-type quarks is redundant. The ``amplification'' occurs 
as large ($\sim 1/2$), approximately equal real entries in $\mat{U}^{u,d}$, 
which cancel against each other in the small elements of $\mat{V}$, 
are multiplied by small relative phases, leading to small imaginary 
contributions to $\mat{V}$, of the right size for $\delta_{\rm KM}$ to be 
large. The matrices $\mat{U}^{u,d}$ are not required to contain large phases.

The larger mixing angles are not much affected by the introduction of 
$\mat{K}$: they get a contribution of the order of $\zeta^2$. 
However, this contribution will be very significant for $V_{13}$ which 
now inherits a relatively large imaginary part; thus, the structure 
of $\mat{V}$ is unavoidably coupled to \CP\ violation. 
With the initial mass matrices of Eq.~(\ref{eq:dem1}) and taking 
$\mat{K}=\mbox{diag} (1,e^{0.008\,i},1)$, one obtains 
$\delta_{\rm KM}=-0.719$ and $J=2.9\times 10^{-5}$, with $|V_{13}|=0.0039$.
Thus, realistic values of \CP-violating parameters can originate from
a small perturbation in conjunction with good values of masses and mixings.

However, it is difficult to imagine the origin of the phases $\zeta 
\vp_i$, amounting to different, generation-dependent phase redefinitions 
for the up and down weak doublet quarks, within our proposal of ``additive'' 
\CP\ violation. 
Also, the degree of amplification (order 100 or less) falls short of that 
achieved by random small imaginary parts. 
This is because the decoupling approach cannot take advantage of the 
amplification of imaginary parts that occurs in the diagonalisation of
the mass matrices, as described in section \ref{sec:diag}. 

\subsection{Weak coupling}\noindent
A different approach is to weakly couple the mass spectrum to \CP\ 
violation. This can be achieved, for example, by choosing the small 
parameters $b$ and $c$ of Eq.~(\ref{eq:dem1}) to be complex \cite{SilvaM01}, 
writing $b e^{i\beta}$ and $c e^{i\gamma}$ instead of $b$ and $c$. 
Although expressed in terms of complex phases, clearly this {\em ansatz}\/ 
does not conflict with Eq.~(\ref{eq:additive}) as long as Im$\,b$, $c$ are 
sufficiently small. 
The invariants of $\mat{H}$ then remain of the same order of magnitude 
but receive small corrections from the phases (corrections would also
be small if Re$\,b$, $c$ were held constant and small imaginary parts 
were added).
In \cite{SilvaM01} all parameters were taken real positive except $c^d$, 
which was assigned a phase $\pi /3$ (on top of its negative sign), 
generating $\delta_{\rm KM}=0.5$ while preserving acceptable values of quark 
masses. The amplification factor $F_{\CP}$ is then found as 
$3\delta_{\rm KM}/(\pi {\rm Im}\, c_d) \simeq m_b/9\sqrt{m_dm_s} \simeq 19$, 
similar to the values resulting from random phases in the down sector and 
hence less favourable for the proposal of small phases, but with the 
advantage of stable $m_d$.

If we choose instead $c^u$ to have a phase $-\pi /3$ then one can  
obtain $\delta_{\rm KM}\simeq -0.5$, with $J= 1.9\times 10^{-5}$, for 
an amplification of $F_{\CP}\sim m_t/9\sqrt{m_um_c} \simeq 600$, with an 
up mass of $3.6\,$GeV (for the same input value $m^u_1$). We see as 
expected that the up sector dominates if one introduces imaginary parts of 
equal magnitude, since the amplification depends on mass ratios.
One may also give $b^u$ an imaginary part of order $|c^u|$, but this is 
less successful, resulting in $J\sim 10^{-6}$ and small $\delta_{\rm KM}$ 
($<0.1$). Such an imaginary perturbation can be partially removed by
phase redefinitions, due to the permutation symmetry: as we will verify
by changing into the ``heavy'' basis, the contribution of Im$\,b$ to the 
perturbations $\zeta_{12}^\h$ and $\zeta_{13}^\h$, for which appreciable 
amplification occur, cancel. It was already shown in \cite{SilvaM98} that 
one would require $b$ to have a large phase ({\em i.e.}\/\ 
Im$\,b\sim{\rm Re}\,b$) in order to generate large $\delta_{\rm KM}$.

Similarly in the three-parameter {\em ansatz}\/ Eq.\ (\ref{eq:demmass}),
to preserve a small up mass one may give small imaginary parts to the 
parameters $B$, $C$, $D$, (keeping either Re$\,B$ or $|B|$, {\em etc.}\/, 
constant).\footnote{In the down sector, the largest allowed phases of 
$B$, $C$ would be about $0.1$, since the mass ratios (and resulting 
small mixing angles) are about $1/30$.}
The most effective way is to give $D^u$ a phase $\pi/3$, which leads to
$\delta_{\rm KM}=-1.13$ and $J=1.9\times 10^{-5}$. The imaginary 
perturbation to $\mat{\Delta}/3$ is $2\,{\rm Im}\,D^u/\sqrt{3} \simeq 
\sqrt{m_u m_c}/m_t$, 
thus the amplification factor is $\delta_{\rm KM} m_t/3\sqrt{m_u m_c} 
\simeq 2000$. Recall that in the democratic basis we are adding phases 
$\pm\sqrt{3}\,{\rm Im}\, D^u$ and it is {\em relative}\/ phases, or
equivalently nonremovable imaginary parts, that induce \CP\ violation).

The effects of a correlated structure of phases on $\det \mat{M}^{u,d}$, 
the up mass and $|V_{13}|$ can be understood intuitively if we change 
to the ``heavy'' basis 
where the initial real matrices are (neglecting $m_i/m_{i+1}$ next to 1)
\beq \label{eq:m0hier}
\mat{m}_0^{({\rm h})} =
\!\left(\ba{ccc} 
0 & \mp \sqrt{m_1m_2} & \pm \frac{\sqrt{m_1m_2}}{\sqrt{2}}\! \\

\!\mp \sqrt{m_1m_2} & -m_2 & -\sqrt{2}m_2 \pm\! \sqrt{\frac{3m_1m_2}{2}}\! \\

\pm \frac{\sqrt{m_1m_2}}{\sqrt{2}} & 
\!-\sqrt{2}m_2 \pm\! \sqrt{\frac{3m_1m_2}{2}}\! & m_3 
\ea \right)\!,\ 
m_3
\!\left(\!\ba{ccc} 0& D& 0\\ D& C& B\\ 0& B& 1
\ea \!\right)
\eeq
for the two- and three-parameter matrices Eq.~(\ref{eq:dem1}) and 
(\ref{eq:demmass}) respectively, and the imaginary perturbation is
\beq \label{eq:zetah}
\zeta \mat{D}^{({\rm h})} = \frac{m_3}{18}\,\cdot \hspace{3.5in}
\eeq
\[ \left(\! \ba{ccc}
3(
-\zeta_{12}-\zeta_{21}) &
\sum \zeta_{i\leftrightarrow j} & 
\!\sum \zeta_{i\leftrightarrow j}\\ 

\!\!\sqrt{3}(
\zeta_{21}\!-\!\zeta_{12}\!
-\!2\zeta_{31}\!+\!2\zeta_{32})& 

\!\zeta_{12}\!+\zeta_{21}\!
-\! 2(\zeta_{31}\!+\!\zeta_{32}\!+\!\zeta_{13}\!+\!\zeta_{23})\!
&
\!\sum \zeta_{i\leftrightarrow j}\\ 

\!\sqrt{6}(
\zeta_{21}\!+\zeta_{31}\!-\zeta_{12}\!
-\zeta_{32})& 

\!\sqrt{2}\left(
\zeta_{12}\!+\zeta_{21}\!+\!\zeta_{31}\!+\!\zeta_{32}\!-\!2(\zeta_{13}\!+\zeta_{23}\!
)
\right)&
\!\mathcal{O}(\zeta) 

\ea\!\right) 
\]
where the notation $\sum \zeta_{i\leftrightarrow j}$ in the $(1,2)$, $(1,3)$, 
$(2,3)$ positions means copy the $(2,1)$, $(3,1)$, $(3,2)$ entry respectively 
but transpose the labels on each of the $\zeta_{ij}$.\footnote{We neglect 
$\zeta_{ij}$ next to 1, and in order to fit the matrix onto the page.} 
We set $\zeta_{11}=\zeta_{22}=\zeta_{33}=0$ by a phase redefinition, without 
loss of generality. Then setting $\zeta_i=0$, but allowing the input values 
of $m_2/m_3$ and $\sqrt{m_1m_2}/m_3$ (or $b$ and $c$, or $B$, $C$ and $D$) 
to be complex in these expressions, will not have a large effect on the size 
of the small mass eigenvalues and mixing angles. Alternatively, one 
can also keep $\mat{m}_0^{({\rm h})}$ real but consider different 
correlated patterns of $\zeta_{ij}$. Clearly adding a small imaginary part 
to $b^u$, equivalent to setting all $\zeta_{i3}$, $\zeta_{3i}$ equal, cannot 
produce sufficient amplification since the crucial $12$ and $13$ elements 
in the heavy basis do not receive imaginary parts (due to cancellations). 
For complex $c^u$, however, effectively written as 
$\zeta_{23}=\zeta_{33}=\zeta_{32}$ (other $\zeta_{ij}$ vanishing), 
the only cancellation is in the $(2,2)$ element.

The ``weak coupling'' proposal, for complex $c$ or $D$ respectively, is 
successful in generating \CP\ violation from very small (order $10^{-3}$) 
imaginary parts, but from the point of view of a theory written in the 
democratic basis, the correlations required between different entries may be 
somewhat contrived. Finally in this section we consider the effect of small 
imaginary perturbations to individual mass matrix elements in the democratic 
basis (other elements being held real).
Then for $m_u$ to remain small, we enforce $\zeta_{ij}=0$, $i,j=1,2$ for the
imaginary perturbation in the democratic basis, and in order for the 
perturbation to affect the small $(1,2)$, $(1,3)$ {\em etc.}\/\ elements of 
$\mat{m}^{({\rm h})}$ and hence potentially produce $J\simeq 10^{-5}$ we 
consider nonzero values of $\zeta^u_{13}$, $\zeta^u_{23}$, $\zeta^u_{31}$, 
$\zeta^u_{32}$ in turn. In the three-parameter mass {\em ansatz}, such a 
perturbation corresponds to a nonzero value of the (1-3) and (3-1) elements of 
$\mat{m}^{({\rm h})}$, which is desirable since recent data on $|V_{13}|$ 
disfavour the exact zero \cite{RRRV-S}.

For $\zeta^u_{13}=-10^{-3}$ we find $\delta_{\rm KM}=-0.82$, $J=2.0\times 
10^{-5}$ but the up mass is marginally affected: $m_u\simeq 6\,$MeV. 
For $\zeta^u_{23}= 10^{-3}$ the results are very similar. For 
$\zeta^u_{31}$ and $\zeta^u_{32}$ of the same size there is very little 
amplification and $\delta_{\rm KM}$ is about $0.0035$, but the overlarge 
value of $m_u$ persists. This difference simply reflects the fact that the 
matrix diagonalising $\mat{m}^u$ on the left, which will determine the 
observable \CP\ violation, ``feels'' two perturbations related by 
transposition differently. The larger value of $m_u$,
compared to the results of taking complex $c^u$ or $D^u$, arises because 
adding imaginary parts to a fixed $\mat{m}_0^{({\rm h})}$ will increase 
the absolute value of the small parameters on which $\det\mat{m}$ and the 
up mass depend. When $c^u$ or $D^u$ receive a phase, the real parts of the 
small parameters of $\mat{m}^{({\rm h})}$ decrease such that $\det\mat{m}$, 
and $m_u$, remain of the same magnitude.


\subsection{Magnification through $V_{13}$}\noindent
If \CP\ violation is to come from a small imaginary perturbation, 
which {\em a priori}\/ is random, {\em i.e.}\/\ does not correspond to any 
particular pattern, then from our initial analysis it follows that it also 
has to be very small, so as not to affect the smallest quark mass 
eigenvalues. However, then $\delta_{\rm KM}$ will be too small, unless 
there is a mechanism that will magnify the influence of some tiny random 
imaginary perturbation, by a factor {\em parametrically larger}\/ than the 
$F_{CP}\gtrsim 500$ (or $\sim 50$ for the down sector) which occurs 
generically in democratic scenarios. 
Clearly, this magnification process has to occur in $V_{13}$ because the 
other off-diagonal elements are too large to be given non-negligible 
phases; $V_{13}$ has to be already very small or even zero in an initial 
real quark mass matrix if the imaginary perturbation is to play a r\^ole. 
In the ``heavy'' basis, 
we require the structure
\begin{equation} \label{eq:V13heavy}
\label{mheavy}\mat{m}^{({\rm h})}_0 = m_3
\left(\begin{array}{ccc}
0 & p & pq \\
p & r & -q \\
pq & -q & 1
\end{array} \right)
\end{equation}
where $q$, $r=\mathcal{O}(m_2/m_3)$ and $p=\mathcal{O}(\sqrt{m_1m_2}/m_3)$. 
The third eigenvalue of the matrix is approximately equal to $m_3$, hence 
we find that the $(13)$ element of the diagonalisation matrix 
$\mat{U}$ is of order $pqr\sim \sqrt{m_1m_2^5/m_3^6} \leq 10^{-5}$. 
As a result the contribution from $\mat{U}^{d}$ to $V_{13}$ is very small. 
(For $\mat{U}^u$ one should look at the (31) element, which is generically 
small anyway.)
But the contribution from $U^{u*}_{21} U^d_{23}$ is likely to be of order 
$(p^u/r^u) q^d = \sqrt{m_u/m_c}\,m_s/m_b \simeq 1.5 \times 10^{-3}$,
so one cannot possibly achieve an ``amplification'' greater than about 
$600$ relative to a small imaginary perturbation in the heavy basis, thus 
random perturbations in the up sector are ruled out as the source of
the observed effect. Since $|V_{13}|$ has now been measured 
to be greater than or equal to $0.003$, (see {\em e.g.}\/\ \cite{RRRV-S}) 
the smallest imaginary part\,\footnote{In a basis where the large entries of 
$\mat{V}$ are real to good approximation.} consistent with large 
$\delta_{\rm KM}$ is order $2\times 10^{-3}$, thus one expects an 
imaginary part of order Im$\,m^{d\h}_{13}\sim 2\times 10^{-3}m_b$ to 
be sufficient. But, on inspecting Eq.~(\ref{eq:zetah}), it is clear 
that random imaginary parts of size $m_b\zeta_{ij}/3$ in the democratic 
basis will give 
rise to Im$\,m^{d\h}_{13}/m_b\sim \zeta_{ij}/6$, since one is adding 
four uncorrelated imaginary parts and the sum is likely to be twice
the average size of each one. The average size of imaginary parts, in 
the notation of section \ref{sec:random}, is $\zeta/2$, thus 
Im$\,m^{d\h}_{13}$, and Im$\,V_{13}$, are expected to be of order 
$\zeta/12$ and the smallest viable value of $\zeta$ is $0.024$. 
Comparing this with the bound on random phases for $m_d$ to be stable, 
$\zeta\leq 12 m_d/m_b \sim 1.5\times 10^{-2}$, the scenario is 
marginally ruled out, but remains the most attractive way of generating
\CP\ violation with small phases in the down sector. If phases smaller 
than $0.024$ in the democratic basis happened to add constructively 
to give larger Im$\,m^{d\h}_{13}$, the scenario could be successful, but 
the probability of such constructive interference is small given random
perturbations $\zeta_{ij}$.
The main disadvantage of the proposal of this section is that it appears 
highly contrived to generate an initial mass matrix in the democratic basis 
which reproduces Eq.~(\ref{eq:V13heavy}) with reasonable accuracy.


In summary, the ``weak coupling'' proposal implemented by complex $b$, 
$c$ or complex $B$, $C$, $D$ in the two mass {\em ans{\" a}tze}\/ 
Eq.~(\ref{eq:dem1},\ref{eq:demmass}), can explain \CP\ violation as a 
small perturbation consistent with the observed masses and mixings, and in 
the following we will take the ``weak coupling'' mass matrices as the 
main examples.

\section{Soft terms}\noindent
The result of applying the proposal of ``additive \CP\ violation'' to 
the relevant soft breaking parameters, the gaugino masses $M_i$, 
scalar bilinear $B$-term and trilinear $A$-terms, is simply to set all 
potentially \CP- and flavour-violating quantities to be of the form 
$\tilde{m}(1+\eps_i+i\zeta_i)$, where $\tilde{m}$ is the magnitude 
predicted by one's favourite mechanism of SUSY-breaking, $|\eps_i|$ are
flavour-dependent parameters, and $|\zeta_i| \lesssim 10^{-3}$. In 
democratic models of flavour, this form may be determined by the 
transformation of the soft terms under $P_L\times P_R\times CP$. The soft
scalar masses transform under either $P_L$ {\em or}\/ $P_R$, hence in the 
limit of unbroken symmetry they take the form
\[ m^2_{0ij} = m_0^2 (\delta_{ij} + \kappa \Delta_{ij}) = m_0^2 
(\delta_{ij} + \kappa), \]
where $i$, $j$ are flavour indices, {\em i.e.}\/\ both the unit and 
democratic matrices are allowed. However, we note that soft scalar 
masses generally turn out to be diagonal in the interaction basis, 
irrespective of the SUSY-breaking mechanism, thus for the time being 
we take $\kappa=0$. Since flavour and \CP\ symmetries are broken, we allow 
deviations from universality, which one might expect to be of order 
${\rm few}\times m_i/m_{i+1}$ or smaller. The flavour breaking 
parameter $m_d/m_s \simeq 0.05\simeq \lambda^2$ is not particularly  
small, thus the SUSY flavour problem is unlikely to be solved without 
a more concrete model relating SUSY-breaking to the flavour symmetry, 
possibly analogously to alignment 
\cite{Nir:1993mx}
for the case of continuous Abelian symmetry.

The $A$-terms are usually written as $\mathcal{L}_{\rm soft}\supset 
-A^u_{ij}Y^u_{ij} Q_i U^c_j H_u + 
(u\rightarrow d) + \cdots$
with flavour symmetry acting the same way on the scalar partners 
as on the fermions. Thus, in democratic models, the couplings 
$\hat{A}^u_{ij}\equiv A^u_{ij}Y^u_{ij}$ (no sum) are expected 
to take the form $A^u_0(1+\eps^{\prime u}_{ij}+i\zeta^{\prime u}_{ij})
y_t/3$, where $A^u_0$ specifies the overall magnitude. The 
coefficients $\eps^{\prime u}_{ij}$, $\zeta^{\prime u}$
may be different from those appearing in $\mat{M}^u$, but are 
expected to be of the same order of magnitude as the coefficients 
$\eps$, $\zeta$ in the Yukawas.\footnote{For example, some complex 
scalar v.e.v.'s may break flavour and \CP, in which case there may 
be complex $F$-terms associated with the same multiplets, but
both the scalar and $F$ components should break the symmetries by 
small amounts.}
Then we may also write (with a similar expression in the down sector)
\beq \label{eq:Aterms}
A^u_{ij}=A^u_0(1+\tilde{\eps}^u_{ij}+i\tilde{\zeta}^u_{ij})
\eeq 
with the coefficients $\tilde{\eps}$, $\tilde{\zeta}$ satisfying the 
same conditions as $\eps'$, $\zeta'$ up to numerical factors of order 1. 
In an explicit model of flavour and SUSY-breaking there may be 
correlations between the parameters $\eps_{ij}$, $\zeta_{ij}$ in the 
Yukawa couplings and $\eps'_{ij}$, $\zeta'_{ij}$ in the $A$-terms, but 
initially we consider a general structure of perturbations.

Gaugino masses and the $\mu$ and $B\mu$ terms are flavour singlets, hence
only transform under \CP: the small perturbation away from a \CP\
invariant theory simply means that the (nonremovable) phases are order 
$10^{-3}$ or less, consistent with Eq.~(\ref{eq:additive}). 

\subsection{Supersymmetric \CP: no longer a problem?}\noindent
Predictions of EDM's depend on the imaginary parts of the rephasing
invariant combinations $\mathcal{M}_A\equiv \hat{A}_{(q,l)}M_i^*$, 
$\mathcal{M}_B \equiv (B\mu)\mu^*M_i^*$, $i=1,2,3$, in the 
quark or lepton mass basis. In the presence of nonuniversal soft terms, 
the effects of \CP- and flavour-violating interactions are most easily 
estimated by changing to the SCKM basis.
The scalar (superpartner) mass matrices then receive off-diagonal or 
imaginary contributions which are treated perturbatively as mass insertions. 
For the EDM calculations one is only interested in the diagonal 
$A$-terms which are usually written $A_u$, {\em etc.}, and enter into 
the observable phases as above.
For a superpartner spectrum not too far above current experimental 
limits, and assuming no correlations between soft term phases, the 
tightest bounds apply to the ``mu phase'' Arg$\,\mathcal{M_B}$
\cite{SUSYEDM}
which is bounded to be $< 10^{-2}$; the $A$-term phases are 
somewhat less restricted with bounds of order $< 10^{-1}$.\footnote{These order-of-magnitude bounds are taken to apply at the 
electroweak scale and are for $|A|$ comparable to soft scalar masses. 
As discussed earlier, bounds on the phases of $\mathcal{M}_{A,B}$ 
only make sense if the absolute magnitudes are specified.}
This is to be compared with the amplification parameters $F_{CP}$ in the 
up and down sectors, of order $\gtrsim 500$ and $\sim 50$ respectively. 
We can easily allow all phases of soft terms to be of order 
$\delta_{\rm KM}/F^u_{CP} < \mbox{few}\times 10^{-3}$, but phases of 
order $1/50\sim\delta_{\rm KM}/F^d_{CP}$ are potentially problematic. 
However, imaginary parts in the down sector need not be order 
$m_d/m_s$, and it is consistent with our realization of approximate \CP\
by small imaginary parts, to set them also to $\mathcal{O}(10^{-3})$.
Thus as far as ``flavour-diagonal'' sources of \CP\ violation are concerned, 
{\em there is no SUSY \CP\ problem}. Assuming that phases of this size are 
present, EDM's should be detected given a moderate improvement 
in experimental sensitivity.

%

\subsection{Nonuniversal $A$-terms and EDM's}\label{sec:nonU}\noindent
If $A$-terms are nonuniversal, then there is a danger that large 
imaginary parts for the (11) entries $\hat{\mat{A}}_{11}^{\rm SCKM}$ 
in the super-CKM basis may be generated, even in the case of soft terms 
which are real in the interaction basis since these entries are no 
longer suppressed by the lightest quark or lepton mass (see 
\cite{stringCP}) but can get contributions proportional to $m_{t,b}$. We 
have
\beq
\hat{\mat{A}}^\dem \rightarrow \hat{\mat{A}}^{\rm SCKM} = 
\mat{U}_L^\dag \hat{\mat{A}}^\dem \mat{U}_R. 
\eeq 
Contributions to $\eps'$ and $\eps$ in the kaon system and $a_{J/\Psi K_S}$
would also be expected through off-diagonal $A$-terms in the SCKM basis.

If our proposal is implemented in the democratic basis, the left and 
right diagonalising matrices are close to the matrix $\mat{F}$ which 
diagonalises $\Delta$. The phases of $\mat{U}_L$ and $\mat{U}_R$ 
need not be large, but
they necessarily contain one or more entries with imaginary part 
$\geq 3 \times 10^{-3}$. For the most favoured ``weak coupling'' 
structures we find $|{\rm Im}\,\mat{U}^{u}_L|$ is order $5 \times 10^{-2}$ 
or less and $|{\rm Im}\,\mat{U}^{d}_L|$ is order $5 \times 10^{-3}$ 
or less. 

To estimate the effects of nonuniversal soft terms we write the 
$A$-terms in the democratic basis as
\[ \hat{A}^{u,d}_{ij} = A_0\left(\mathbf{1}+
{\eps}^{u,d}(\mat{D}^{u,d}_0+ \tilde{\mat{D}}^{u,d}_0) 
+i{\zeta}^{u,d}(\mat{D}^{u,d}+\tilde{\mat{D}}^{u,d})
\right)_{ij} \frac{y_{t,b}}{3}, \]
where, on the expectation that the flavour structure in the soft terms
will be parallel to that of the Yukawas, we take 
$\tilde{D}_{0ij}$, $\tilde{D}_{ij}$ to be order 1 (where $\eps$, $\zeta$, 
$\mat{D}_0$, $\mat{D}$ are as defined in Eq.~(\ref{eq:epszeta}) and
${\eps}^u\sim 0.015$, $\tilde{\eps}^d\sim 0.1$, ${\zeta}^{u,d}\sim 10^{-3}$).
It is convenient to first change to the heavy basis 
\beq
\hat{\mat{A}}\rightarrow \hat{\mat{A}}^{({\rm h})} = 
\mat{F} \hat{\mat{A}} \mat{F}^\dag 
\eeq 
in which, in the absence of any correlated structure for the 
$\tilde{D}_{(0)ij}$, we expect the nonuniversal contributions
also to be of order $A_0 (\tilde{\eps}^{u,d}+i\tilde{\zeta}^{u,d}) 
y_{t,b}/3$.
In this basis also, the imaginary parts of the diagonalisation matrices 
are order $5 \times 10^{-2}$ $(5 \times 10^{-3})$ or smaller in the up (down) 
sector; we already know that the real parts are given by diagonalising 
Eq.~(\ref{eq:m0hier}).
Then the mass insertions Im$\,(m^2_{11})_{LR}$ in the super-CKM basis 
arise from multiplying $\hat{A}$ by the appropriate Higgs v.e.v.:
\bea \label{eq:LRmasses}
{\rm Im}\, (m^2_{11})_{LR} &\simeq& - m_3 A_0 \Biggl(
\zeta{\rm Re}\,U^{({\rm h})\dag}_{1iL} \frac{\tilde{D}^{({\rm h})}_{ij}}{3} 
{\rm Re}\,U^{({\rm h})}_{j1R} + \nonumber \\ 
&+& \eps \Bigl( {\rm Im}\,U^{({\rm h})\dag}_{1iL} \frac{\tilde{D_0}^{({\rm h})}_{ij}}{3}{\rm Re}\,U^{({\rm h})}_{j1R} + {\rm Re}\,U^{({\rm h})\dag}_{1iL} \frac{\tilde{D_0}^{({\rm h})}_{ij}}{3}
{\rm Im}\,U^{({\rm h})}_{j1R}
 \Bigr) \Biggr)
\eea
in either the up or down sector, where the first term arises from 
complex $A$-terms in the theory basis, and the other terms come from 
nonuniversality of (real) $A$-terms.\footnote{Of course Im$\,A_0$ will also
contribute, but this ``universal'' contribution was already considered in the
previous section and does not dominate over the contributions considered 
here.} Considering the first term only, one can take $i=j=1$ to find a 
contribution giving rise to Im$\,(m^2_{11})^u_{LR}/\tilde{m}^2 \sim 
10^{-3}A_0m_t/3\tilde{m}^2 \sim 10^{-4}$, 
which is unlikely to be cancelled by any other term to good enough 
accuracy to respect the EDM bounds (of order $10^{-6}$--$10^{-7}$). 
Thus even without considering the ``string CP'' contributions from
real nonuniversal A-terms, uncorrelated small imaginary parts for 
$A_{ij}$ are ruled out. This problem, arising from the $(11)$ element 
in the heavy basis, is analogous to the up mass problem in the case of 
the Yukawas, which suggests that similar non-generic structures of 
imaginary parts in the $A$-terms may help to evade the EDM bounds 
while still allowing some nonuniversality. To take a simple example, if 
$A$-terms are of the form $\hat{\mat{A}} = \mat{A}_L\cdot 
\mat{y}+\mat{y}\cdot \mat{A}_R$, where $\mat{A}_{L,R}$ are diagonal 
matrices \cite{KobayashiV}, the (11) elements in the SCKM basis are 
still proportional to $m_{u}$ or $m_d$, thus the EDM's are as small as 
for universal $A$-terms (given small imaginary parts of $\mat{A}_{L,R}$).

\subsection{Nonuniversal benchmarks with additive CP violation}
We now investigate whether some other restricted structures of 
nonuniversal A-terms, correlated to the Yukawa matrices, could also 
lead to suppressed EDM's. 
For each of the quark mass matrices 
Eq.~(\ref{eq:dem1},\ref{eq:demmass}), we find the consequences 
if the $A$-terms have the same flavour structure 
but with different coefficients, which we will call $\tilde{b}$, 
$\tilde{c}$, or $\tilde{B}$, $\tilde{C}$, $\tilde{D}$, respectively.
This form has not been derived from an underlying theory, but is 
imposed as a reasonable starting point or benchmark, based on 
Eq.~(\ref{eq:additive}) and on the expectation that the same operators
generating quark flavour will also produce nonuniversality. 
If this {\em ansatz}\/ turns out to be ruled out, even this restricted 
form of nonuniversality cannot be allowed; if it is successful, it 
motivates a search for theories in which such structures appear, and may 
suggest some characteristic signals of new physics. For example, if the 
quark mass matrix is as in Eq.~(\ref{eq:dem1}) we take
\beq \label{eq:demsusy1}
\mat{A}^{u,d} = \frac{A_0m_3}{3}\left( \ba{lll}
1 & 1 & 1+\tilde{b} \\
1 & 1 & 1+\tilde{b}-\tilde{c} \\
1+\tilde{b} & 1+\tilde{b}-\tilde{c} & 1+\tilde{b}-\tilde{c}
\ea \right)
\eeq
with real parts of $\tilde{b}$ and $\tilde{c}$ randomly chosen on 
$9m_2/2m_3(-1,1)$ and $3\sqrt{3m_1m_2}/m_3$$(-1,1)$ respectively. Note
that while we used the running quark masses defined at a scale of 
$1\,$GeV in the analysis of section \ref{sec:demo} ({\em e.g.}\/\ we used
$m_t(1\,{\rm GeV})\simeq 400\,$GeV), the appropriate RG scale is now 
the electroweak scale $M_Z$, thus the prefactor $m_3(M_Z)$ will be 
smaller, approximately by a factor 2. 
For the quark matrices of Eq.~(\ref{eq:demmass}) one follows an exactly
analogous procedure.
Since these forms are merely a starting point for evaluating nonuniversal 
contributions, we do not consider RG running from high energies. 

For each quark mass matrix {\em ansatz}\/ we consider two scenarios: one 
``conservative'', in which only $c^{u,d}$ and $\tilde{c}^{u,d}$, or 
$D^{u,d}$ and $\tilde{D}^{u,d}$ are complex, and one ``less conservative'' 
in which $b^{u,d}$, or $B^{u,d}$ and $C^{u,d}$, and the corresponding 
tilded $A$-term parameters are complex ({\em i.e.}\/\ all small parameters 
are allowed to be complex). Imaginary parts are at most of order Im$\,c^u$, 
or Im$\,D^u$, respectively.

Taking $\tilde{m}=400\,$GeV, we obtain for the mass matrices 
Eq.~(\ref{eq:dem1}), in the ``conservative'' case, that the typical
values of Im$\,(\delta^u_{11})_{LR}$ and Im$\,(\delta^d_{11})_{LR}$ are 
order $5\times 10^{-6}$, $5\times 10^{-7}$ respectively. Thus the 
contribution of $A_u$ to the EDM's is marginally too 
large; however, one requires only mild cancellation (at the 
$10\%$ level) to respect the bound. Accidental cancellations might 
occur between different contributions to $A_u$: for example, we find 
that the first term involving $\zeta\tilde{\mat{D}}^u$ in 
Eq.~(\ref{eq:LRmasses}), which arises from Im$\,\tilde{c}^u$, 
is of the same order as the cross-term 
Im$\,U^{u\dag}_{L1j}{\rm Re}\,\tilde{c}^u$, and may be of opposite sign. 
In the ``less conservative'' case, results are very similar, indicating
that the $b$ and $\tilde{b}$ parameters have little influence on the EDM's.
Other imaginary parts of mass insertion parameters $(\delta_{ij})_{LR}$
are of order $10^{-4}$ or smaller in the up sector and $10^{-6}$ in the
down, hence contributions to flavour-changing CP-odd observables appear to 
be negligible. Real parts of off-diagonal $(\delta_{ij})_{LR}$ are also 
below experimental FCNC bounds \cite{Gabbiani96}, thus our approach is
self-consistent, in that the form of nonuniversality that we have chosen
is not ruled out by such bounds before even considering EDM's.

For the mass matrices Eq.~(\ref{eq:dem1}), in the ``conservative'' case,
we find that Im$\,(\delta^u_{11})_{LR}$ vanishes (or is order $10^{-8}$
if $A_0$ has a small overall phase) while Im$\,(\delta^d_{11})_{LR}$ is 
order $10^{-6}$ or smaller; in the ``less conservative'' case, both
Im$\,(\delta^u_{11})_{LR}$ and Im$\,(\delta^d_{11})_{LR}$ are order $10^{-6}$ 
or smaller. Thus in this {\em ansatz}\/, which has slightly smaller 
coefficients of flavour symmetry-breaking operators, it appears easier to 
satisfy the EDM bounds with minimal fine-tuning or cancellations, 
consistent with a certain degree of nonuniversality.

\section{Summary}\noindent
If the world is approximately supersymmetric, it may be possible to test 
theories of the origin of \CP\ violation and flavour, as well as the 
mechanism of SUSY-breaking, in the near future. However, first one must 
explain the absence, at the level of current experimental sensitivity, of 
EDM's and flavour-changing processes resulting from superpartner loops. 
If the flavour problem is solved by universality or heavy scalars, and the 
\CP\ problem by a mechanism giving automatically real soft terms, then we 
learn little about the origin of \CP\ violation and flavour, which may lie
at an arbitrarily high scale; conversely, if general soft terms are 
allowed then the parameter space is too large and the experimental 
constraints too complex for meaningful investigation. 

We propose a guiding principle, {\em additive \CP\ violation}, that
provides a realisation of \CP\ violation as a small perturbation consistent
with experimental data, allowing a large CKM phase and nonzero phases of 
soft terms. In essence \CP\ is broken by adding small imaginary parts 
to real couplings (rather than, for instance, by large phase rotations of 
real couplings). Applying this principle to universal soft terms, EDM's 
are predicted just below the experimental bounds; for nonuniversal $A$-terms 
we require particular structures of deviations from universality to satisfy 
the bounds. Such structures are no more fine-tuned than the small 
perturbations to Yukawa couplings, which we know are required to generate 
quark masses and mixings, so one might reasonably expect whatever mechanism 
leads to Yukawa structure to also generate correlated patterns of soft terms. 
%

An essential part of the proposal is the {\em amplification}\/ which 
automatically generates the observed size of $\delta_{\rm KM}$ and $J$ 
from small phases in the democratic basis, or equivalently small 
imaginary parts of Yukawa couplings in any flavour basis: one may have  
Im$\,m_{ij}\leq 3\times 10^{-4}m_t$ in the up sector and still obtain  
$\delta_{\rm KM} \sim 0.5$. Amplification of Im$\,\mat{m}^{u,d}$ occurs 
in the diagonalisation of quark mass matrices, and small imaginary parts 
in the $\mat{U}^{u,d}$ matrices can easily produce large $\delta_{\rm KM}$.
There are several ways to implement \CP\ violation as a small imaginary 
perturbation while ensuring correct masses and mixings, the most attractive
being to link the perturbation to a small parameter breaking flavour 
symmetry.

The Yukawas and soft terms that we use are not derived from a fundamental
theory, nevertheless they can be seen as a consequence of flavour and 
\CP\ symmetries broken by small parameters. The motivation for our 
phenomenological investigation is to draw attention to the possibility that 
complex and even non-universal soft terms are allowed, without requiring 
unnatural fine-tuning compared to the Yukawa couplings, and to provide 
guidelines for future model-building efforts which may give more definite 
predictions for new physics signals. Independently of the model, the 
parameter breaking \CP\ cannot be smaller than $3\times 10^{-4}$, thus 
nonzero EDM's should be within one or two orders of magnitude of current 
bounds and the scenario may be testable within a few years. 

If signals are found near the current level of sensitivity, the 
interpretation is either that the \CP-violating parameter is somewhat
larger, or that there is a mild degree of nonuniversality which enhances
the $A$-term contributions.

\section*{Acknowledgments}
We acknowledge useful conversations with Gordy Kane, Ikaros Bigi, 
R.\,M.~Mohapatra, Anna Teixeira and Shaaban Khalil. Research supported in 
part by DOE Grant DE-FG02-95ER40899 Task G.


\end{document}